\def\year{2016}
\documentclass[letterpaper]{article}
\usepackage{aaai}
\usepackage{times}
\usepackage{helvet}
\usepackage{hhline}
\usepackage{courier}
\usepackage{wrapfig,lipsum,booktabs}
\usepackage{balance}  
\usepackage{graphics} 
\usepackage{txfonts}
\usepackage{times}    
\usepackage[pdftex]{hyperref}
\usepackage{color}
\usepackage{textcomp}
\usepackage{booktabs}
\usepackage{ccicons}
\usepackage{todonotes}
\usepackage{xcolor}

\usepackage{helvet}
\usepackage{courier}
\usepackage{times}
\usepackage{url}
\usepackage{latexsym}
\usepackage{graphicx, color, soul}
\usepackage{balance}
\usepackage[font={small}]{caption}
\usepackage[labelfont=bf]{caption}
\usepackage[]{algorithm2e}
\usepackage{subcaption}
\usepackage{tabularx}
\usepackage{comment}
\usepackage{wrapfig}
\def\mvp{\vspace*{-0.1in}}
\nocopyright

\begin{document}

\title{\textit{Tweeting the Mind and Instagramming the Heart}:\\ Exploring Differentiated Content Sharing on Social Media}
\author{Lydia Manikonda \qquad Venkata Vamsikrishna Meduri \qquad Subbarao Kambhampati \\ Department of Computer Science, Arizona State University, Tempe AZ 85281\\
\{lmanikon, vmeduri, rao\}@asu.edu
}

\maketitle
\begin{abstract}
Understanding the usage of multiple OSNs (Online Social Networks) has been of significant research interest as it helps in identifying the unique and distinguishing trait in each social media platform that contributes to its continued existence. The comparison between the OSNs is insightful when it is done based on the representative majority of the users holding active accounts on all the platforms. In this research, we collected a set of user profiles holding accounts on both Twitter and Instagram, these platforms being of prominence among a majority of users. An extensive textual and visual analysis on the media content posted by these users revealed that both these platforms are indeed perceived differently at a fundamental level with Instagram engaging more of the users' $heart$ and Twitter capturing more of their $mind$. These differences got reflected in almost every microscopic analysis done upon the linguistic, topical and visual aspects.

\end{abstract}
\mvp

\section{Introduction}
\label{sec:intro}

Online Social Networks (OSNs) are gaining attention for being rich sources of information about individuals including many aspects of their daily life through the way they connect, communicate and share information. Over the past few years, given their ubiquity and accessibility, social media platforms like Twitter and Instagram have emerged as very popular microblogging services for web users to communicate with each other through text and photos. In 2015, when Instagram broke the record of having more than 400 million active monthly users, Twitter was projected as its main rival. In fact according to a recent article by Pew Research,\footnote{http://www.pewinternet.org/2015/08/19/the-demographics-of-social-media-users/} 28\% of American adults use Instagram and 23\% use Twitter. More interestingly, many users have active accounts on both these sites (or platforms)~\cite{Lim2015mytweet,ChenKDD14understand}. While research has recognized immense practical value in understanding the user behavioral characteristics on these platforms separately, there is no existing research that has examined \textit{how the content posted by individuals differs across these two platforms}. Instagram is a photo-sharing application whereas Twitter emerged as a text-based application which currently lets users post both text and multimedia data. Of particular interest is the question of \textit{why and how individuals use these two sites when both of them are similar in their current functionalities.}

In this research, we aim to answer the aforementioned questions by analyzing content from the same set of individuals across these two popular platforms and quantifying their posting patterns (we focus on ordinary users who are common users but not celebrities or popular users or organizations). By leveraging NLP and Computer Vision techniques, we present some of the first qualitative insights about the types of trending topics, and social engagement of the user posts across these two platforms. Analysis on the visual and linguistic cues indicates the dominance of personal and social aspects on Instagram and news, opinions and work-related aspects on Twitter. 
Despite considering the same set of users on both platforms, we see remarkably different categories of visual content -- predominantly eight categories on Instagram and four categories of images on Twitter. These results suggest that Instagram is largely a sphere of positive personal and social information where as Twitter is primarily a news sharing media with higher negative emotions shared by users. 

\mvp\mvp
\paragraph{Background:} Twitter has been explored extensively with respect to the content~\cite{Honey2009beyond}, language~\cite{Lichan11lang}, etc and it is established that it is primarily a news medium~\cite{Kwak2010news}. Research on Instagram has focused mostly on understanding the user behavior through analyzing color palettes~\cite{Nadav2012Visualize}, categories~\cite{Yuheng14Web}, filters~\cite{Bakhshi2015Filter}, etc. On the other hand, it has been of significant interest to the researchers to investigate the behavior of a user~\cite{Benevenuto2009user}, connect users~\cite{Zafarani2013users}, study how users reveal their personal information~\cite{Chen2012merrier}, etc all across multiple OSNs. We extend the current state of the art by examining the nature of a given user's behavior manifested across Twitter and Instagram. Close to our work is the work of Bang et al.~\cite{Lim2015mytweet} where six OSNs were studied to analyze the temporal and topical signature (only w.r.t user's profession) of user's sharing behavior but they did not focus on studying the comparative linguistic aspects and visual cues across the platforms. Here we employ both textual and visual techniques to conduct a deeper analysis of content on both Twitter and Instagram.
\mvp\mvp
\paragraph{Dataset:} In order to investigate and characterize a given user's behavior across multiple sites, we use a personal web hosting service called \textit{About.me} (http://about.me/) that enables individuals to create an online identity by letting them provide a brief biography, connections to other individuals and their personal websites. Using its API, we performed the data collection of 10,000 users and pruned the individuals who do not have profiles on both Instagram and Twitter. The final crawl includes 1,035,840 posts from Twitter (using the Twitter API \url{https://dev.twitter.com/overview/api}) and 327,507 posts from Instagram (using the Instagram API \url{https://www.instagram.com/developer/}) for the same set of users. Each post in this dataset is public and the data include user profiles along with their followers and friends list, tweets (insta posts), meta data for tweets that include favorites (likes), retweets (Instagram has no explicit reshares; so we use comments in lieu of the attention the post receives), geo-location tagged, date posted, media content attached and hashtags.

\mvp\mvp
\section{Text Analysis}

\subsection{Latent Topic Analysis}
\label{sec:LDA}
In order to explain the types of content posted by a user across Twitter and Instagram, we first mine the latent topics from the corpus of Twitter (aggregated posts on Twitter of all users) and corpus of Instagram (aggregated posts on Instagram of all users where we use captions associated with posts for this analysis). 
We use TwitterLDA (\url{https://github.com/minghui/Twitter-LDA}) developed for topic modeling of short text corpora to mine the latent topics. With the user accounts obtained from \textit{About.me}, the topic inference is meaningful as it is pertinent to the bi-platform posts from users who use both the social media venues.

\begin{figure}
\tiny
\begin{center}
\includegraphics[width=0.35\textwidth]{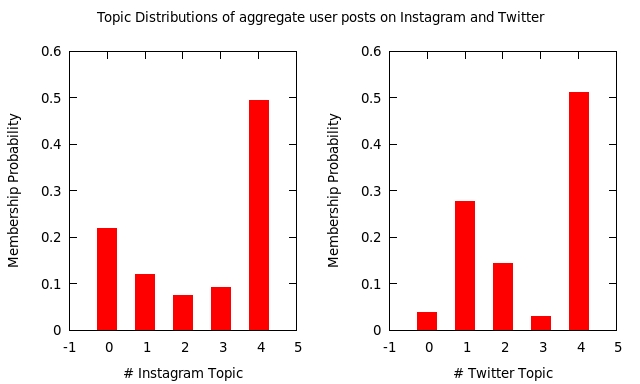}
\caption{Topic distributions of all the user posts on Twitter and Instagram}
\label{fig:aggTwitInstaDist}
\vspace{-0.1in}
\end{center}
\end{figure}

The topic vocabulary listed for both the platforms in Table~\ref{tab:ldatopics} indicates the unique topics for each site as well as the overlapping topics. For instance topics 0 and 4 on Instagram are similar to the topics 1 and 2 on Twitter. However, a significant difference is that Instagram is predominantly used to post about art, food, fitness, fashion, travel, friends and family but Twitter hosts a significantly higher percentage of posts on sports, news and business as compared to other topics. Another notable difference is that the vocabulary from non-English language posts like French and Spanish is higher on Twitter as compared to the captions on Instagram mostly using English as the language medium. The topic distributions obtained from the two corpora are listed in Figure~\ref{fig:aggTwitInstaDist} which show that friends and food are the most frequently posted topics on Instagram as against sports and news followed by work and social life being popular on Twitter. 

\begin{table}
\vspace{-1mm}
\sffamily
\tiny
\centering
\begin{tabular}{| p{3.5cm} | c | p{3.5cm} | } \hline
\textbf{Twitter} & \textbf{ID} & \textbf{Instagram} \\ \hline
stories, international, food, web, n{\~a}o, angelo, j{\'a} & 0 & \#food, delicious, coffee, sunset, beautiful, happy, \#wedding \\ \hline
time, people, love, work, world, social, life & 1 & \#streetart, \#brightongraffiti, \#belize, \#sussex, \#hipstamatic, \#urbanart, \#lawton \\ \hline
happy, love, home, birthday, weekend, beautiful, park & 2 & \#fashion, \#hair, \#makeup, \#health, \#workout, \#vegan, \#fit \\ \hline
m{\'a}s, d{\'i}a, v{\'i}a, gracias, mi, si, las & 3 & \#instagood, \#photooftheday, \#menswear, \#style, \#travel, \#beach, \#summer \\ \hline
\#football, \#sports, \#news, \#art, facebook, google, iphone & 4 & birthday, beautiful, love, christmas, friends, fun, home \\ \hline
\end{tabular} \vspace{-.07in}
\caption{Words corresponding to the 5 latent topics from Twitter and Instagram} \vspace{-.1in}
\label{tab:ldatopics}
\vspace{-2mm}
\end{table}

\begin{figure}
\tiny
	\centering
\mvp\mvp
\mvp\mvp
	\includegraphics[width=0.3\textwidth]{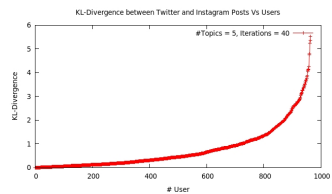}
\mvp\mvp
\mvp\mvp
	\caption{Sorted entropies between the topic distributions of the user posts on Twitter and Instagram}
	\label{fig:entropyVsUsers}
\mvp\mvp
\end{figure}

To further validate the observations made about the distinctive topical content across the two platforms, we compared the topic distributions for each individual on the two platforms by estimating the \textit{KL-Divergence} (entropy) for each user. However for this entropy computation to be possible, a unified topic model needed to be built on the combined corpus of tweets and captions of Instagram posts. 
The unified topics are listed in the description of Figure ~\ref{fig:soceng}. 
The resultant entropy plot in Figure~\ref{fig:entropyVsUsers} follows a power law distribution showing that most users post on Twitter and Instagram equally differently barring a few (where the estimated $p$-value $< {10}^{-15}$ for each user). 


\mvp
\subsection{Social Engagement}
\label{sec:soceng}

\begin{figure}
    \centering
	\tiny
    \begin{subfigure}[b]{0.28\textwidth}
        \includegraphics[width=\textwidth]{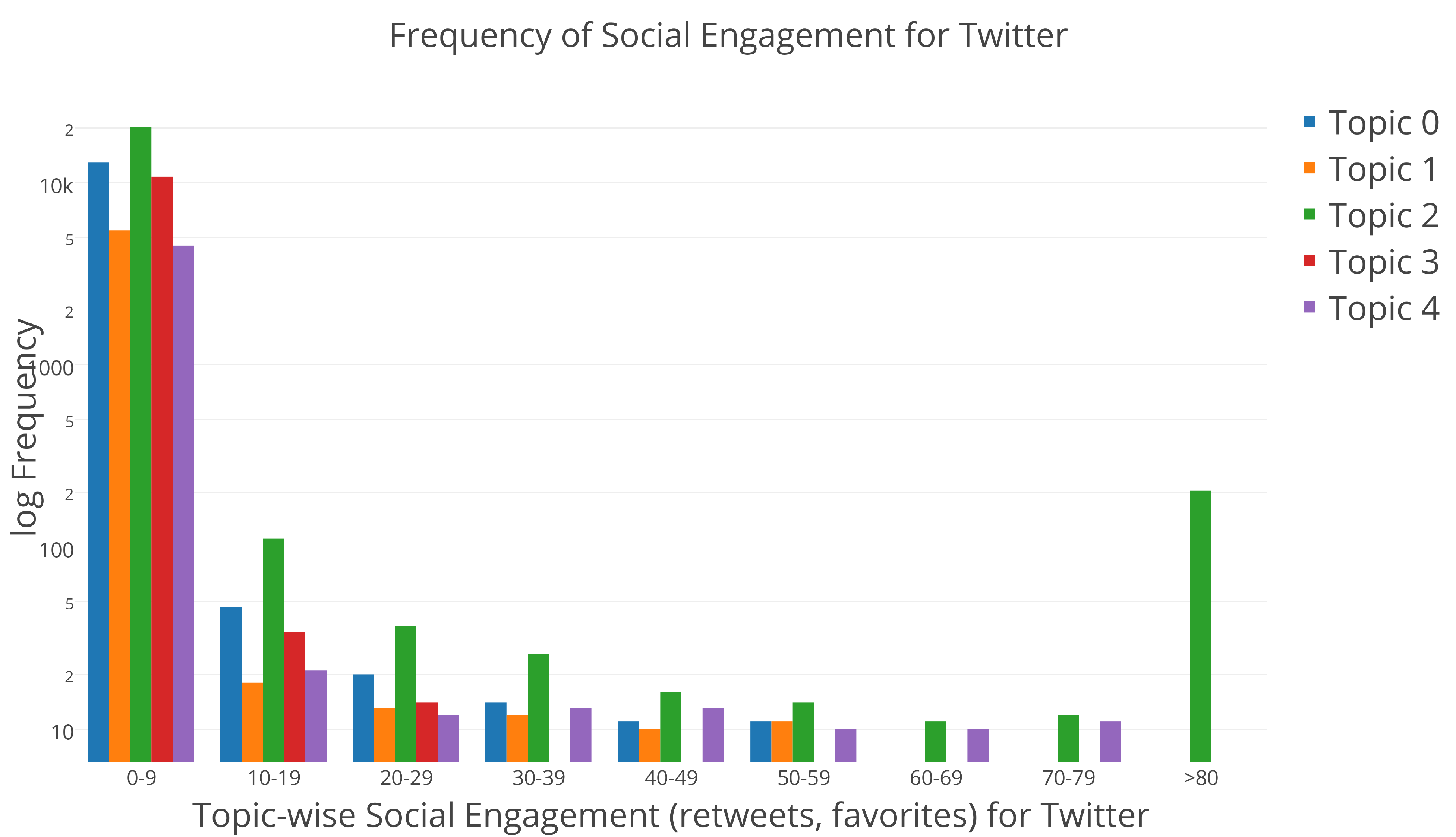}
        \caption{\textit{Twitter}}
    \end{subfigure}
    \begin{subfigure}[b]{0.28\textwidth}
        \includegraphics[width=\textwidth]{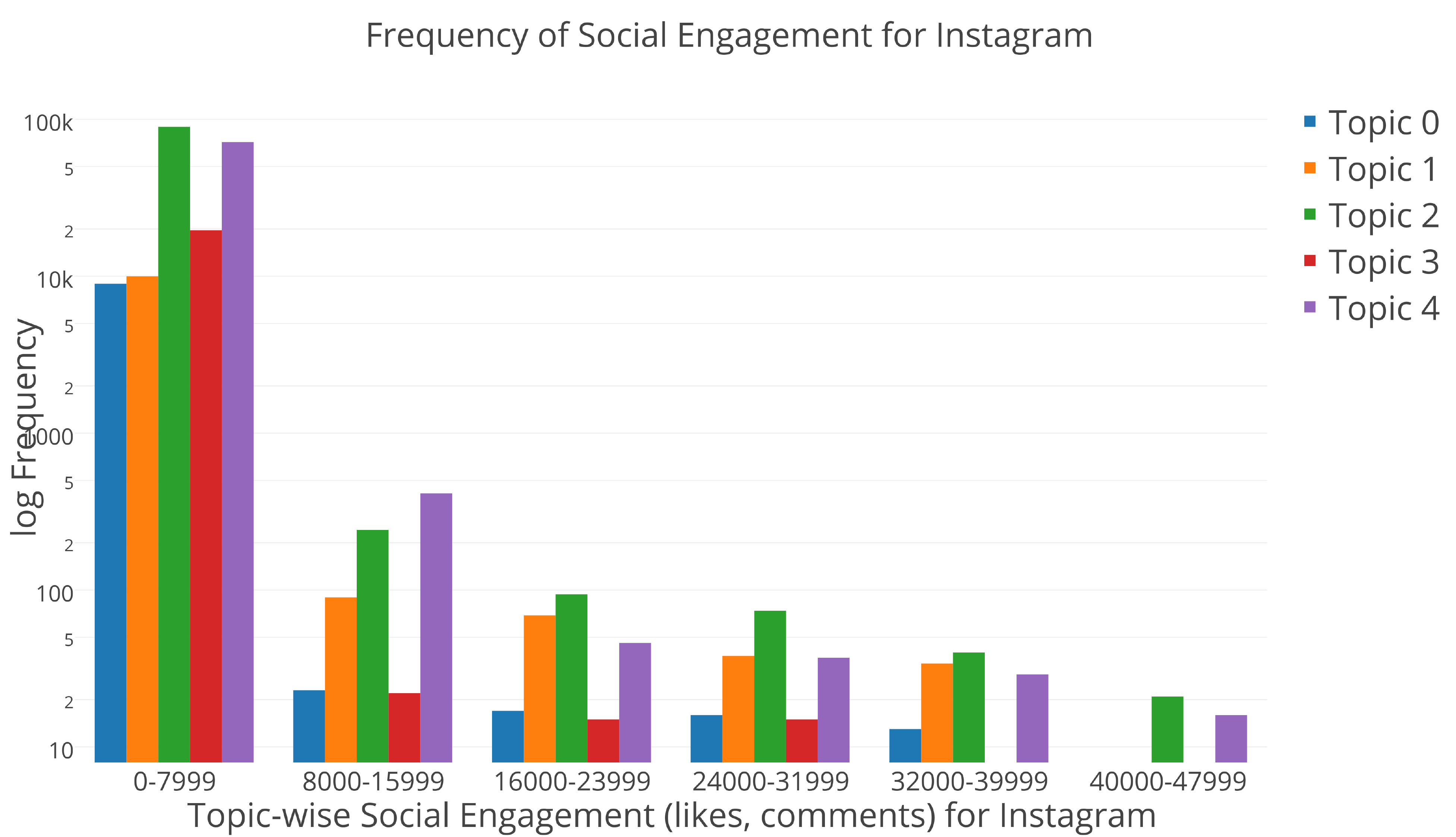}
        \caption{\textit{Instagram}}
    \end{subfigure}
    \vspace{-0.1in} \caption{Social Engagement Vs Post Frequency where the topics are -- Topic 0:\{people, life, world, social, app, game, business\}, Topic 1:\{stories, artists, \#lastfm, level, \#football, \#sports, news\}, Topic 2:\{birthday, beautiful, work, weekend, park, dinner, christmas\}, Topic 3:\{ yang, run, \#fitness, \#runkeeper, \#art, sale, \#menswear\}, Topic 4:\{\#instagood, \#photooftheday, \#love, m{\'a}s, \#fashion, \#travel, \#food\} }
    \label{fig:soceng}
    \vspace{-0.17in}
\end{figure}

Since our findings revealed that the bi-platform topics are significantly different and so we wanted to investigate how these posts made by the same user engage other individuals on the two sites. We define the social engagement as the attention received by a user's post on the social media platform and can be quantified in various ways ranging from the sum of likes and comments on Instagram and the sum of favorites and reshares on Twitter. 
For each topic in the unified topic model for both Twitter and Instagram, the logarithmic frequency of posts is plotted against the magnitude of social engagement that is binned to discrete ranges in Figure~\ref{fig:soceng}. 

An interesting observation is that the socially engaging topics in the combined model are same as the overlapping topics from the topic models built in isolation on the Twitter and Instagram posts (Figure ~\ref{fig:aggTwitInstaDist} in Section ~\ref{sec:LDA}). The dominating topic on Twitter is about sports, news and business but the overlapping influential topic is about social and personal life comprising friends and family. Surprisingly, we found that the overlapping topics (Topics 2 and 3) fetched predominant social engagement on both Twitter and Instagram. 

\begin{figure}[t]
	
    \centering
	\tiny
    \includegraphics[width=0.4\textwidth]{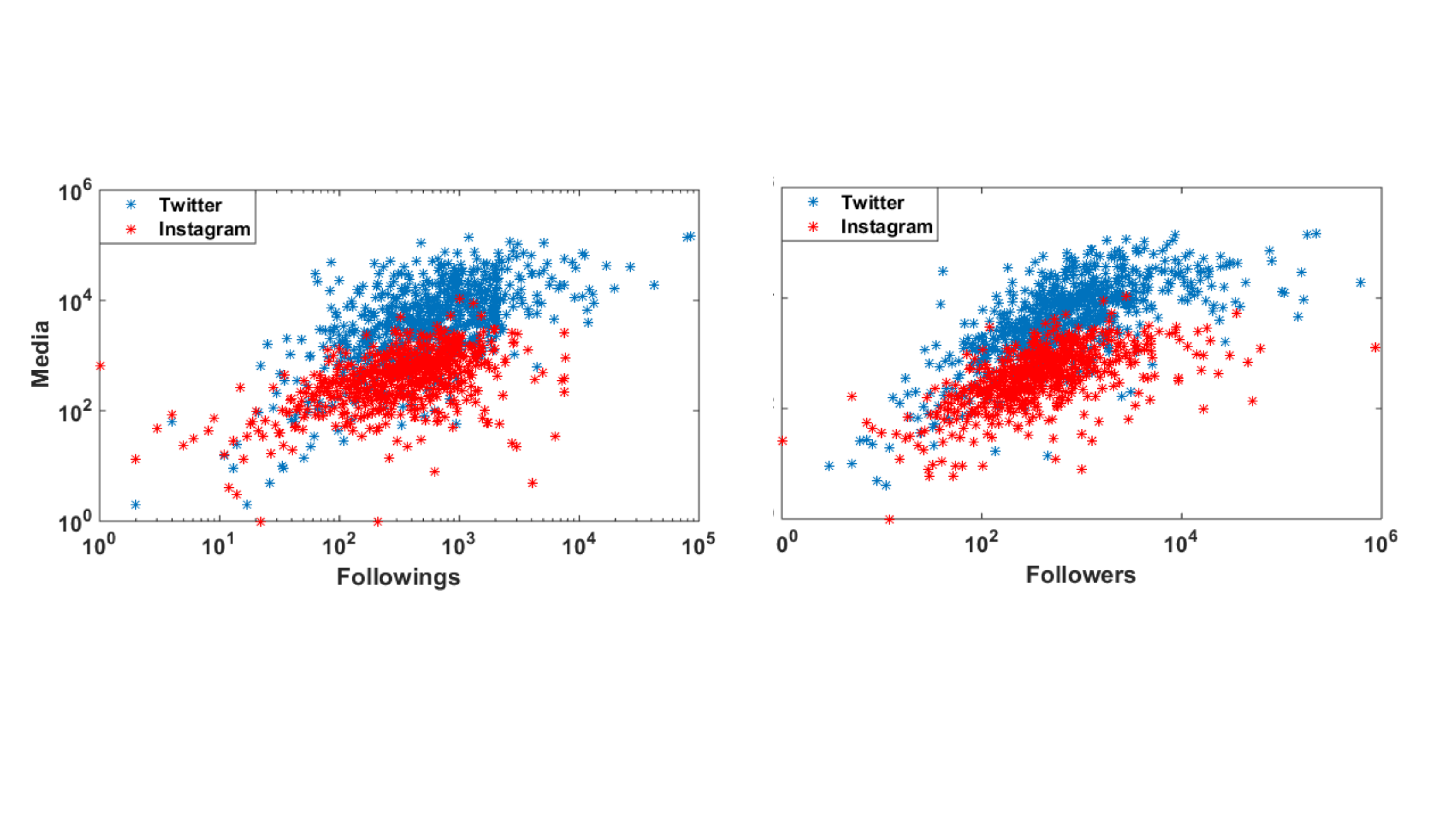}
	\mvp\mvp\mvp\mvp\mvp
    \caption{Distributions of Followers/Followings vs Media}
    \label{fig:follmed} \vspace{-.07in}
	\mvp\mvp
\end{figure}

A notable difference between the platforms with respect to social engagement is that the magnitude of attention received for Instagram posts is significantly higher than the level of attention received on Twitter as we can notice from the ranges plotted on the $x$-axes in Figure~\ref{fig:soceng}. This observation is consistent regardless of the activity of the user. Even when a user is more active (Figure~\ref{fig:follmed}) on Twitter than Instagram, the observation of higher social engagement on Instagram on an absolute scale holds. A possible explanation to this is that the users on Twitter use it as a news source to read informative tweets but not necessarily all of the content that is read will be ``liked''.

On average, there are 30\% more number of hashtags for a Twitter post compared to an Instagram post (Pearson correlation coefficient $= 0.34$ between distributions with $p$-value $< {10}^{-15}$). This may also indicate that on Instagram since the main content is image, textual caption may not receive as much attention from the user.
\mvp
\subsection{Linguistic Nature}
To characterize and compare the type of language used on both platforms, we use the psycholinguistic lexicon LIWC (\url{(http://liwc.wpengine.com/)}) on the text associated with a Twitter post and an Instagram post. We obtain measures of attributes related to user behavior -- \textit{emotionality} (how people are reacting to different events), \textit{social relationships} (friends, family, other humans) and \textit{individual differences} (attributes like bio, gender, age, etc). 

\begin{table}
\sffamily
\tiny
\centering
\begin{tabular}{lcc}
\toprule
& \multicolumn{2}{c}{Platform}\\
\cmidrule{2-3}
& Twitter & Instagram \\ 
\midrule
\textbf{Emotionality} & \multicolumn{2}{c}{}\\
\midrule
Negemo & 0.60 & 0.49 \\
Posemo & 0.19 & 0.19 \\ 
\midrule
\textbf{Social Relationships} & \multicolumn{2}{c}{}\\
\midrule
home & 0.15 & 0.30 \\
family & 0.14 & 0.21 \\
friend & 0.05 & 0.1 \\ 
humans & 0.17 & 0.21 \\
\midrule
\textbf{Individual Differences} & \multicolumn{2}{c}{}\\
\midrule
work & 0.81 & 0.5 \\
bio & 0.6 & 0.93 \\ 
swear & 0.08 & 0.06 \\
death & 0.07 & 0.04 \\
gender & 0.16 & 0.2 \\
\bottomrule
\end{tabular}
\caption{Linguistic attributes across Twitter vs Instagram. Each value indicates the fraction of a post belonging to the corresponding attribute}
\label{tab:llinguistics}
\mvp\mvp
\end{table} 

It is clear from 
Table~\ref{tab:llinguistics} that posts on Twitter are more negatively emotional and contain more work-related and swear words where as positive social patterns are more evident on Instagram. By relating these results to the topic analysis results in the previous section, we identify that on Instagram users share less negative content and more light-hearted happy personal updates. To further support these claims from the textual data, $n$-$gram$ analysis indicates that users on Instagram focus on things that give them pleasure such as, fashion or travel (top bi/Trigrams like \textit{last night}, \textit{Good morning}, \textit{right now}, \textit{fashion design streetwear}), whereas on Twitter they mainly share check-in feeds from their apps or news (top bi/Trigrams like \textit{Stories via}, \textit{Just posted}, \textit{@YouTube video}, \textit{Just posted photo}).

\mvp

\section{Visual Analysis}
\label{sec:visualAnalysis}

Extensive studies have been conducted on the textual and visual content on these two platforms in isolation but to the best of our knowledge, a comparative content analysis has never been conducted. Considering this as a main objective, this section develops a better understanding on the types of photos individuals post on Twitter in comparison with their Instagram posts. To achieve this we employ Computer Vision techniques mainly in terms of -- visual categories (kinds of photos) and visual features (color palettes).  
\mvp
\subsection{Visual Categories}
We first sampled two sets of 5K images from both platforms separately and using the OpenCV library (\url{http://opencv.org/}) on these two datasets, we extracted Speeded Up Robust Features (SURF) for each image. We used the vector quantization approach on these features that eventually converted each image into a codebook format. Using the codebook, we clustered images using $k$-means algorithm (best value of $k$ is found by SSE (Sum of Squared Error)). We employed the same approach on both datasets separately. To our surprise the clusters we obtained on Instagram were very refined compared to the coarse nature of clusters on Twitter dataset. After computing these clusters, two researchers separately identified the overall themes of these two datasets and agreed upon the final visual categories of the photos from these platforms. 

\begin{figure}[t]
   \tiny
    \centering
    ~ 
    \begin{subfigure}[b]{0.11\textwidth}
        \includegraphics[width=\textwidth]{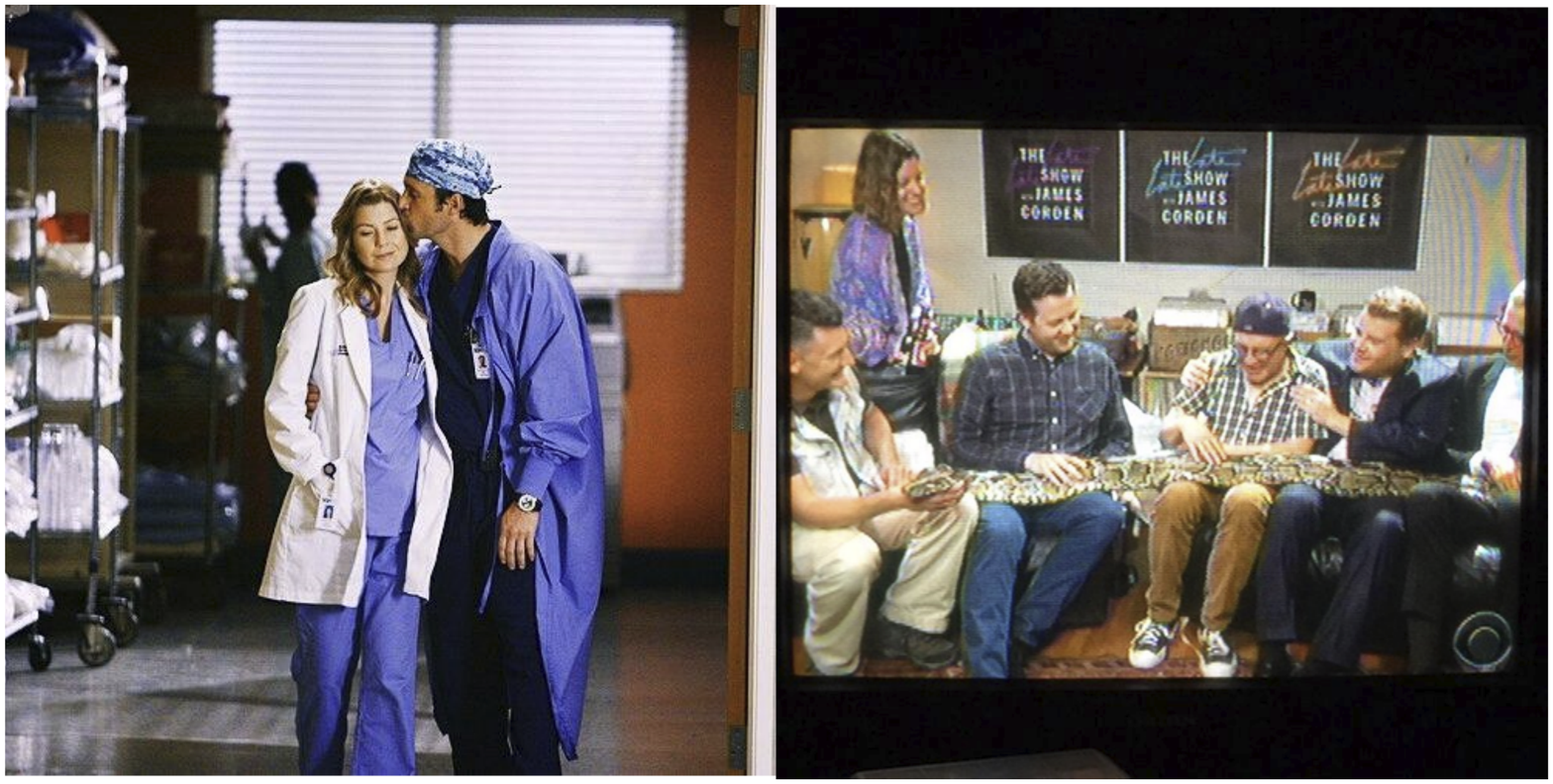}
	\vspace{-0.2in}
        \caption{}
    \end{subfigure}
    ~ 
    \begin{subfigure}[b]{0.11\textwidth}
        \includegraphics[width=\textwidth]{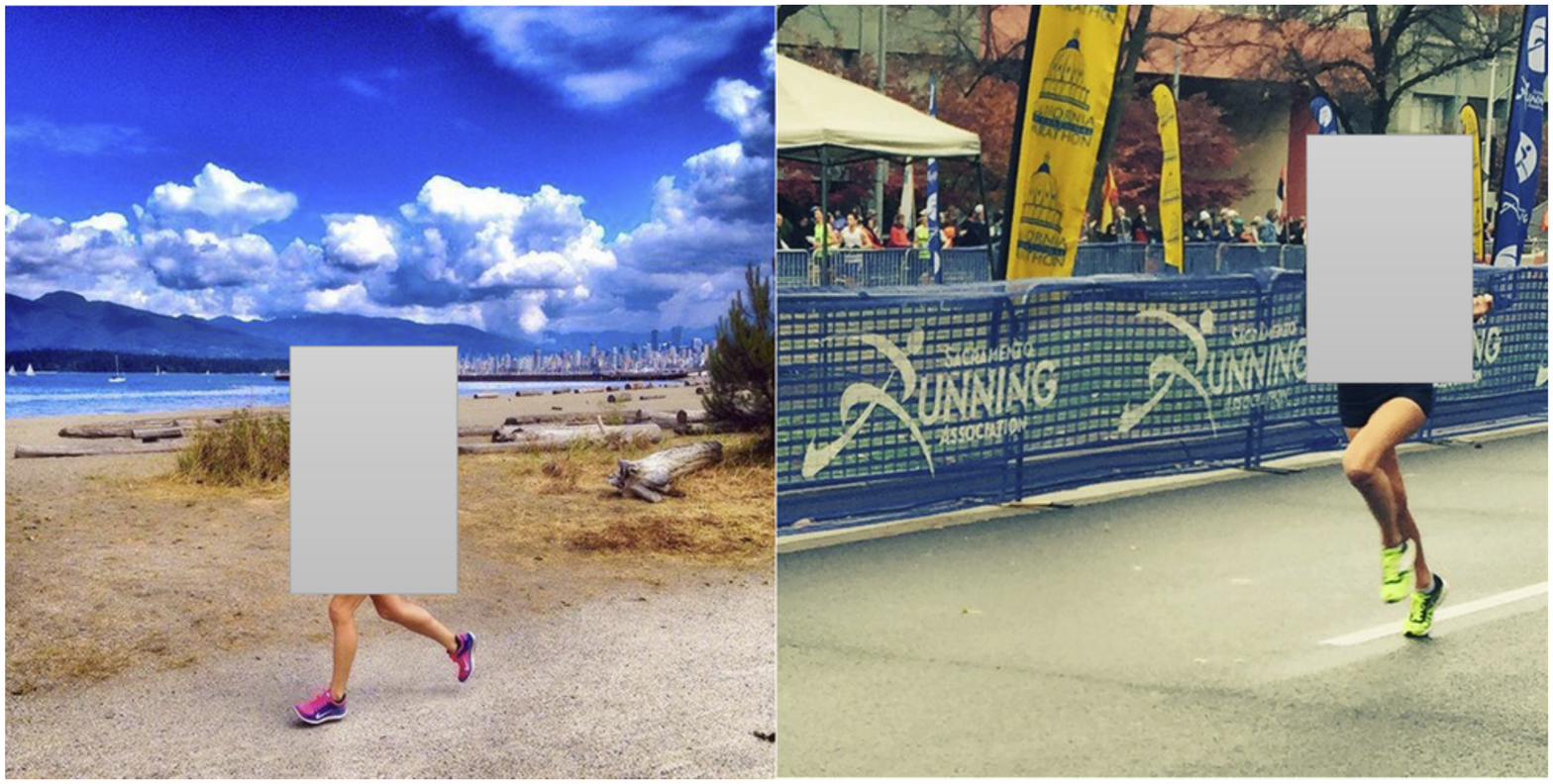}
	\vspace{-0.2in}
        \caption{}
    \end{subfigure}
    ~
    \begin{subfigure}[b]{0.11\textwidth}
        \includegraphics[width=\textwidth]{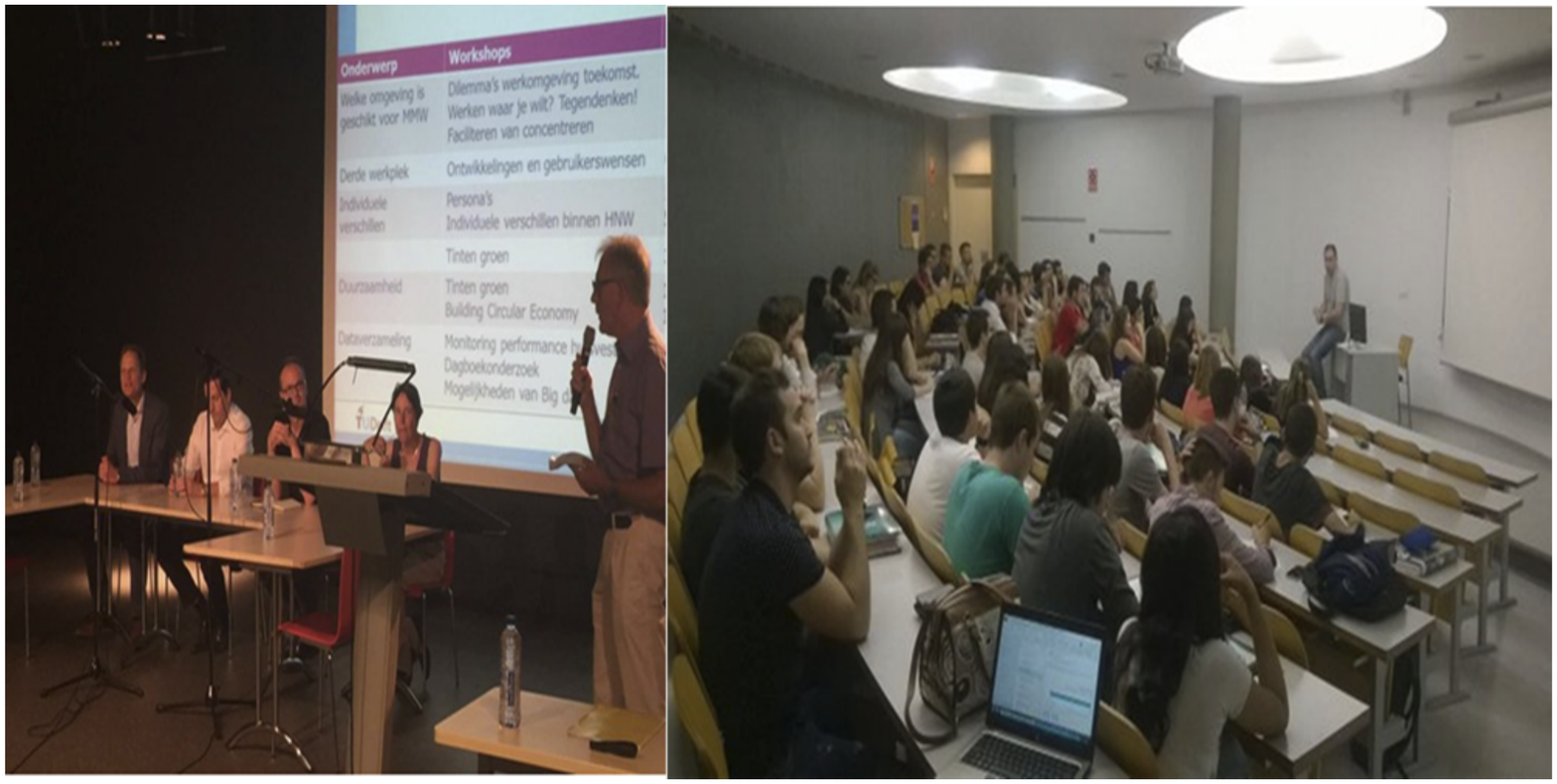}
	\vspace{-0.2in}
        \caption{}
    \end{subfigure} 
    ~
    \begin{subfigure}[b]{0.11\textwidth}
        \includegraphics[width=\textwidth]{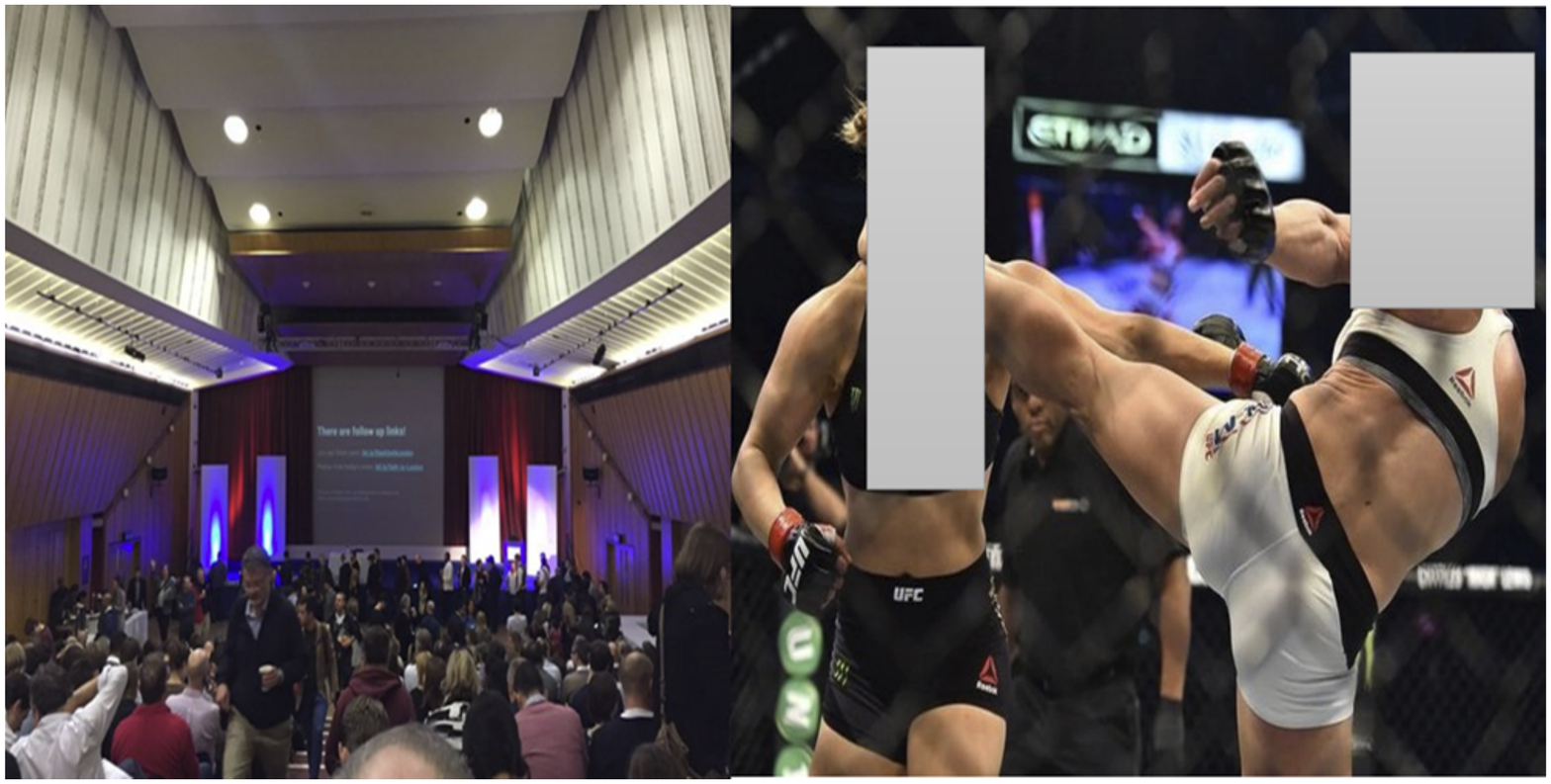}
	\vspace{-0.2in}
        \caption{}
    \end{subfigure}
    \vspace{-0.1in} \caption{Subcategories of \textit{activity}: a) TV shows, b) Running, c) Conferences, d) Live shows}\label{fig:activity}
    \vspace{-0.17in}
\end{figure}

\begin{figure}[t]
    \tiny
    \centering
    ~ 
    \begin{subfigure}[b]{0.11\textwidth}
        \includegraphics[width=\textwidth]{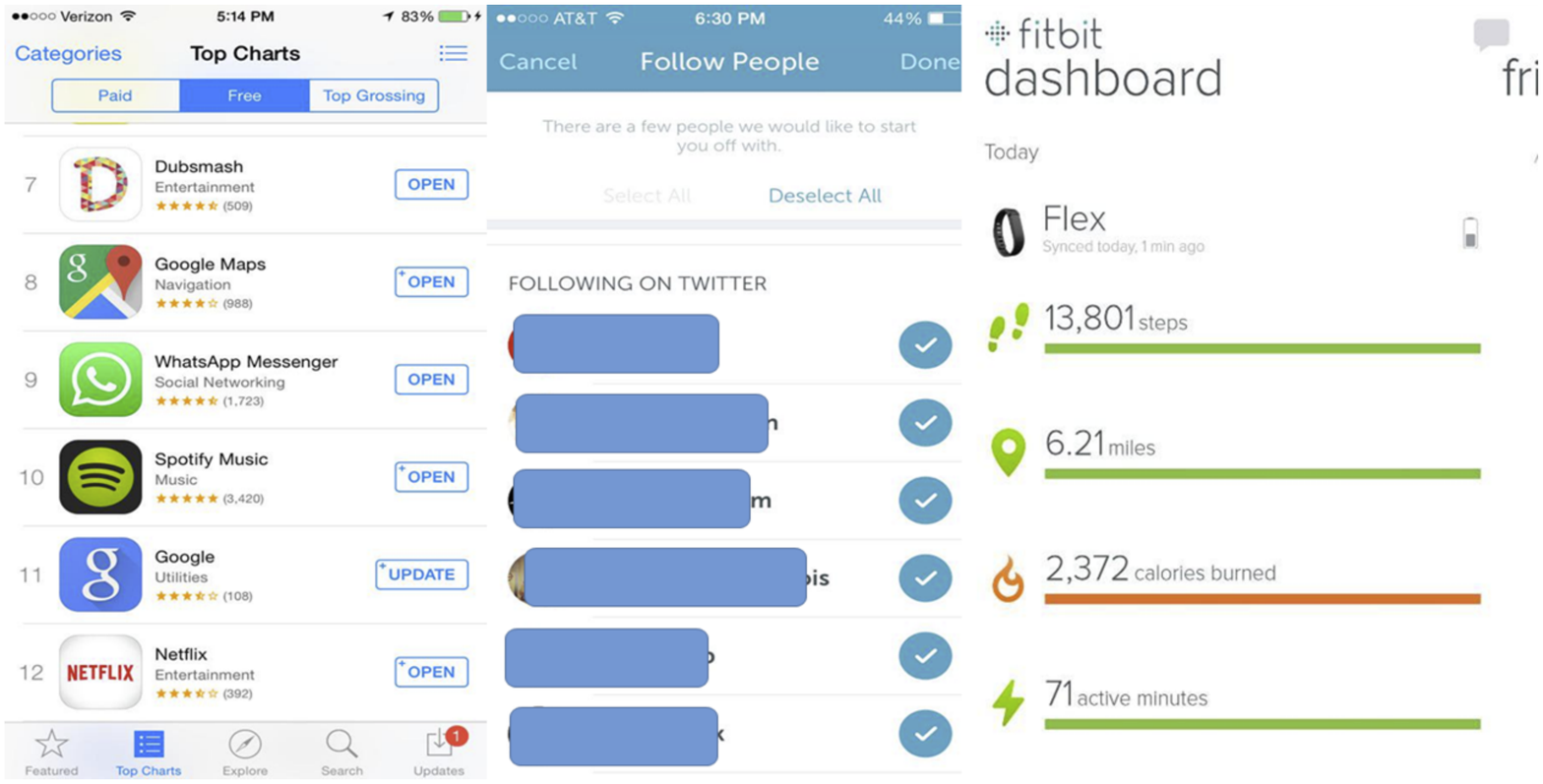}
	\vspace{-0.2in}
        \caption{}
    \end{subfigure}
    ~ 
    \begin{subfigure}[b]{0.11\textwidth}
        \includegraphics[width=\textwidth]{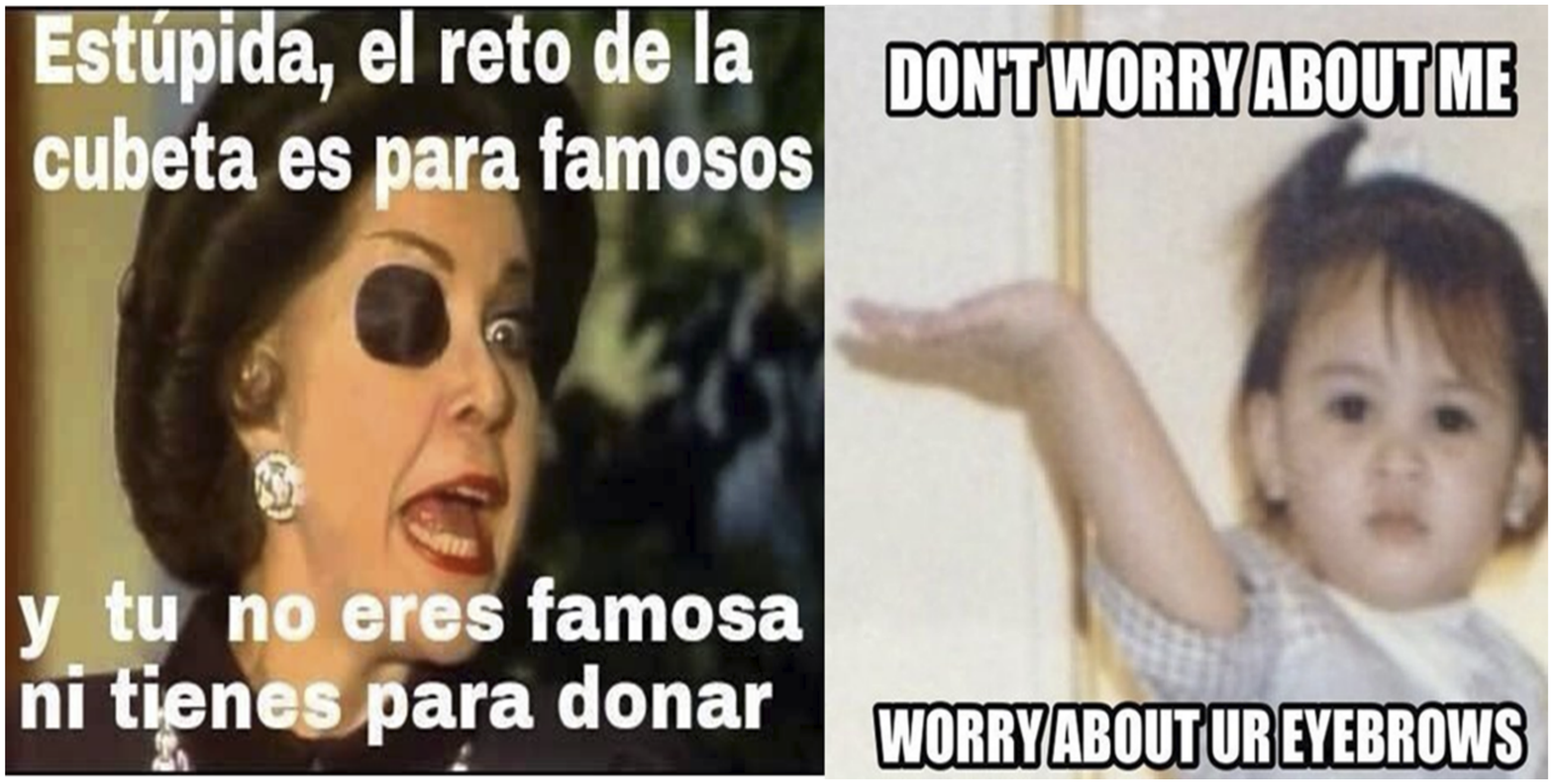}
	\vspace{-0.2in}
        \caption{}
    \end{subfigure}
    ~
    \begin{subfigure}[b]{0.11\textwidth}
        \includegraphics[width=\textwidth]{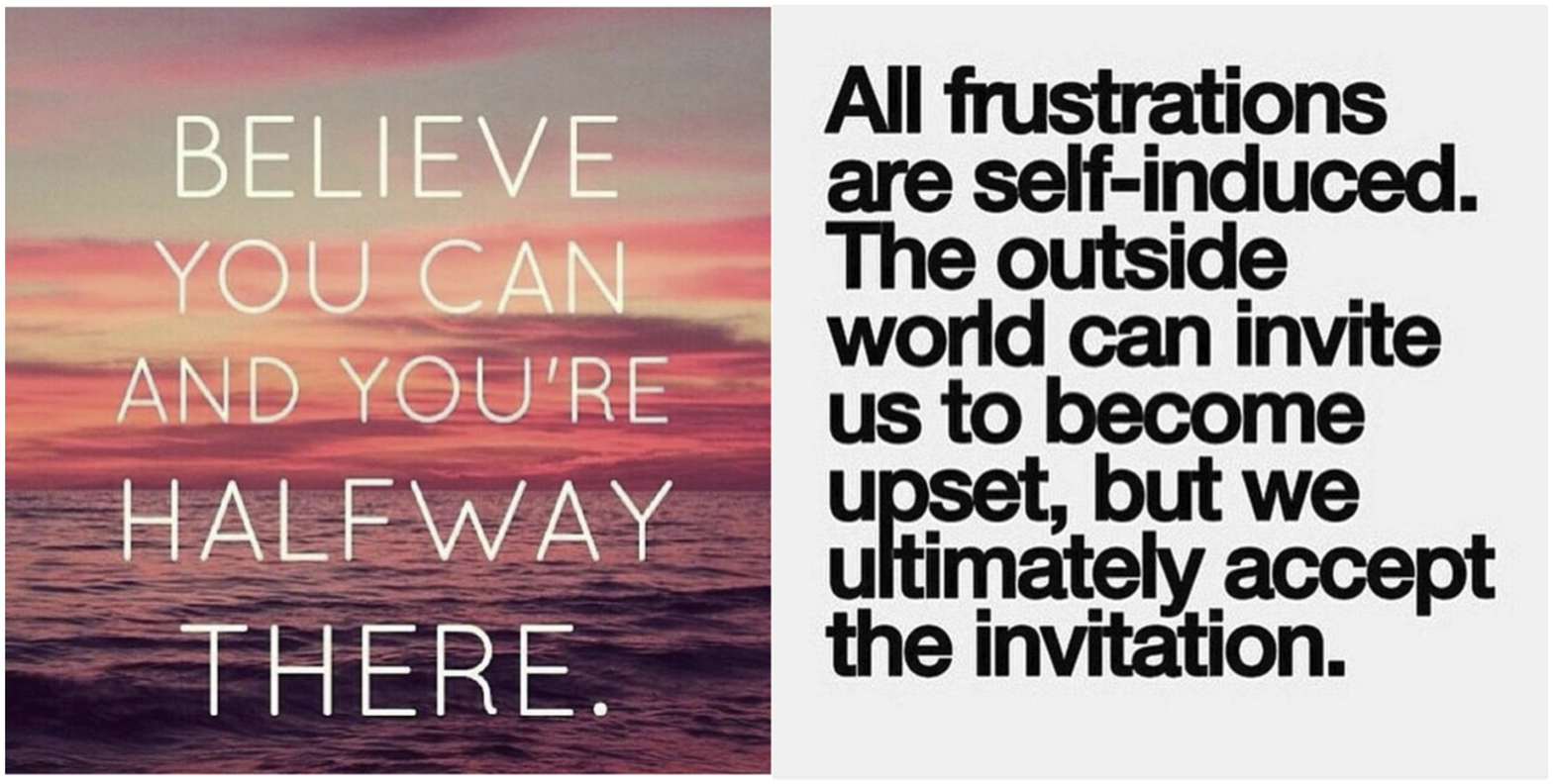}
	\vspace{-0.2in}
        \caption{}
    \end{subfigure} 
    \vspace{-0.1in} \caption{Subcategories of \textit{captioned photos}: a) Snapshots, b) Memes, c) Quotes}\label{fig:capphotos}
    \vspace{-0.17in}
\end{figure}

\begin{figure}[t]
\tiny
    \centering
    ~
    \begin{subfigure}[b]{0.22\textwidth}
        \includegraphics[width=\textwidth]{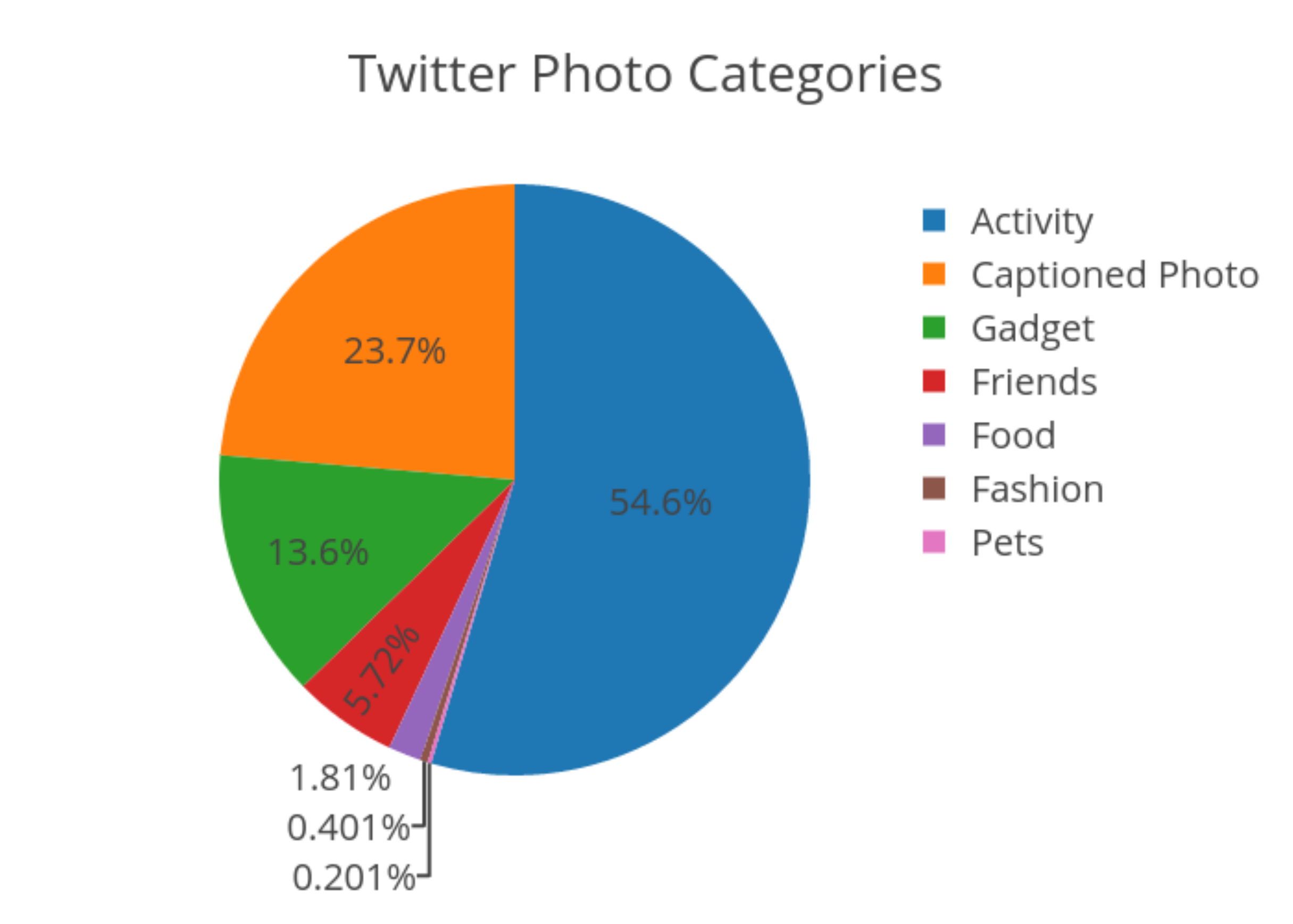}
    \end{subfigure}
    ~ 
    \begin{subfigure}[b]{0.22\textwidth}
        \includegraphics[width=\textwidth]{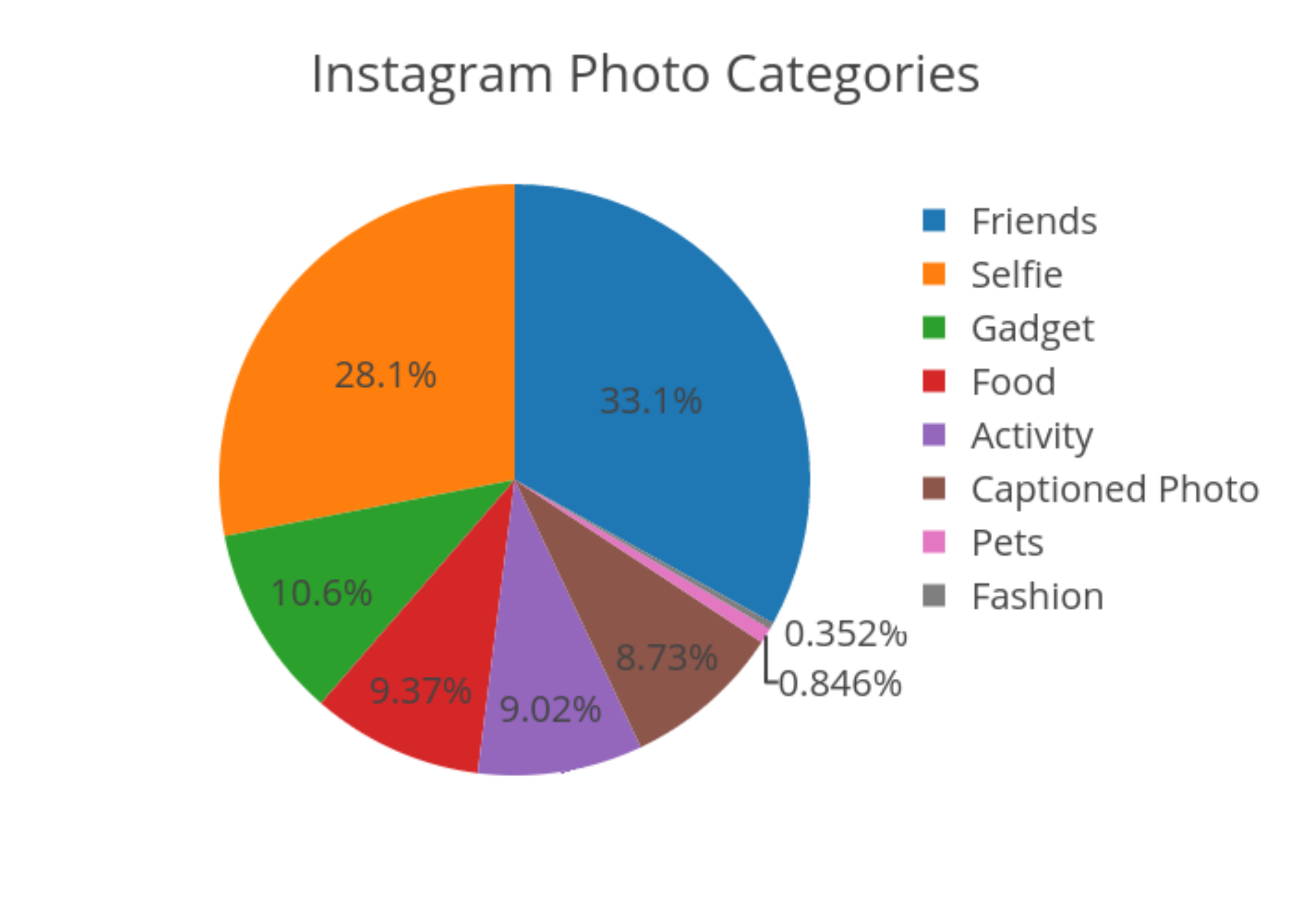}
    \end{subfigure}
    \caption{Photo categories on Twitter vs Instagram}\label{fig:visualcats}
	\mvp
\end{figure}

Visual categories on Instagram agree with our previous work~\cite{Yuheng14Web} which detected eight different categories of images. We tried to categorize the Twitter images into the same format as Instagram images and there are four prominent cluster categories on Twitter. Figure~\ref{fig:visualcats} shows that the percentage of photos in the activity category outnumbered any other category followed by captioned photos. To better understand the kinds of activities and captions shown in these two sections, we sampled around 200 images and asked the two researchers to label them manually into different sub-categories. Figure~\ref{fig:activity} indicates the most popular sub-categories in the \textit{activity} category -- news, events (football games, concerts, conferences) and races and Figure~\ref{fig:capphotos} indicates that majority of the captioned photos are snapshots, memes, and quotes or opinions. These categories suggest  that the topics of photos on Twitter are mainly related to news, opinions or other general user interests where as on Instagram they mainly share their joyful and happy moments of their personal lives.

\mvp
\subsection{Visual Features}
\label{sec:colorHist}
\begin{figure}[t]
\tiny
    \centering
    ~ 
    \begin{subfigure}[b]{0.22\textwidth}
        \includegraphics[width=\textwidth]{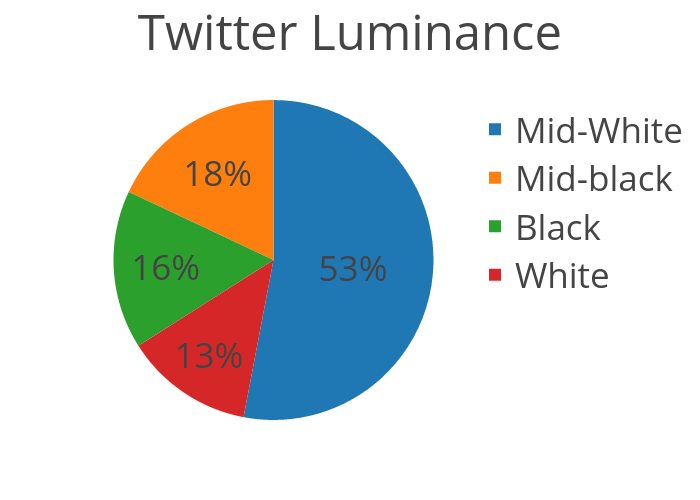}
    \end{subfigure}
    ~ 
    \begin{subfigure}[b]{0.22\textwidth}
        \includegraphics[width=\textwidth]{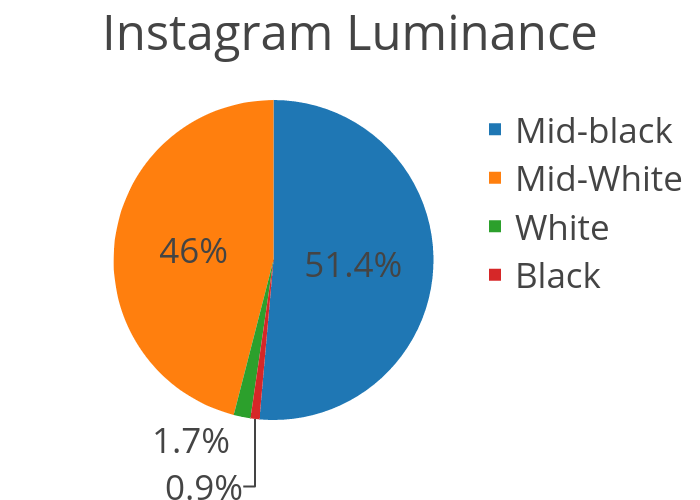}
    \end{subfigure}
    \caption{Example images corresponding to the four major color categories obtained by extracting color histograms of images associated with the Twitter and Instagram posts.}\label{fig:colorhistpics}
	\mvp
\end{figure}

Existing literature~\cite{Bakhshi2015Filter} shows that the images with a single dominant color gain more popularity on Instagram. To verify this we compared the visual luminance of all images on these two platforms. We extracted the grayscale histograms (range from 0 to 259) by utilizing the OpenCV library as they capture the information about the brightness, saturation and contrast distribution. The images with darker pixels were binned into the low intensity value bins close to 0 and images with brighter pixels were binned into the high intensity values close to 259. Later we clustered the images in our dataset by employing $k$-means algorithm with the grayscale histograms as features for each image on both these platforms following which 4 types of clusters were detected based on their color distributions. We measured the image distribution across each of these 4 categories for both the platforms. We noticed that the images on Instagram containing darker and brighter pixels are negligible when compared to Twitter as shown in Figure~\ref{fig:colorhistpics}. This suggest that Twitter posts may be less socially engaging than Instagram owing to a huge presence of captioned photos on Twitter. 


\mvp
\section{Conclusions}

In this paper, we presented a detailed comparison of the textual and visual analysis of the content posted by the same set of users on both Twitter and Instagram. Some of the insights obtained from linguistic analysis reveal the fundamental differences in the thinking style and emotionality of the users on these two platforms and how the posts receive varying degrees of attention as per the underlying topics. Interestingly, user posts on Instagram seem to receive significantly more attention than Twitter. The visual analyses with respect to categories and color palettes indicate that the pictures posted on Instagram contains more selfies and photos with friends where as Twitter contains more about user opinions in the form of captioned photos -- memes, quotes, etc. We observed that the differences are deeply rooted in the very intention with which users post on these platforms with Twitter being a venue for serious posts about news, opinions and business life where as Instagram acting as the host for light-hearted personal moments and posts on leisure activities. 


\balance
\small
\mvp
\bibliographystyle{aaai}
\bibliography{diffusion}

\end{document}